\documentclass[letterpaper]{article}

\usepackage{emulateapj}
\usepackage{onecolfloat}
\usepackage{apjfonts}
\usepackage{amsmath}
\usepackage{natbib}
\usepackage{graphicx}

\newcommand{\Msun}      {\mbox{$\rm\,M_{\mathord\odot}$}}

\submitted{Accepted by the Astrophysical Journal}

\begin{document}

\twocolumn[
\title{X-Ray Flares and Oscillations from the Black Hole Candidate X-Ray 
Transient XTE~J1650--500 at Low Luminosity}

\author{John A. Tomsick\altaffilmark{1},
Emrah Kalemci\altaffilmark{2},
St\'ephane Corbel\altaffilmark{3}, 
Philip Kaaret\altaffilmark{4}}

\altaffiltext{1}{Center for Astrophysics and Space Sciences, Code
0424, University of California at San Diego, La Jolla, CA,
92093 (e-mail: jtomsick@ucsd.edu)}

\altaffiltext{2}{Space Sciences Laboratory, University of California, 
Berkeley, CA, 94702-7450}

\altaffiltext{3}{Universit\'e Paris VII and Service d'Astrophysique, 
CEA Saclay, 91191 Gif sur Yvette, France}

\altaffiltext{4}{Harvard-Smithsonian Center for Astrophysics, 60 Garden Street,
 Cambridge, MA, 02138}

\begin{abstract}

We report on X-ray observations made with the {\em Rossi X-ray Timing
Explorer} of the black hole candidate (BHC) transient XTE J1650--500 
at the end of its first, and currently only, outburst.  By monitoring 
the source at low luminosities over several months, we found 6 bright 
$\sim$100~s X-ray flares and long time scale oscillations of the X-ray 
flux.  The oscillations are aperiodic with a characteristic time scale 
of 14.2 days and an order of magnitude variation in the 2.8-20~keV flux.  
The oscillations may be related to optical ``mini-outbursts'' that have 
been observed at the ends of outbursts for other short orbital period 
BHC transients.  The X-ray flares have durations between 62 and 215~s 
and peak fluxes that are 5-24 times higher than the persistent flux.  
The flares have non-thermal energy spectra and occur when the persistent 
luminosity is near $3\times 10^{34} (d/4~{\rm kpc})^{2}$ erg~s$^{-1}$ 
(2.8-20~keV).  The rise time for the brightest flare demonstrates that 
physical models for BHC systems must be able to account for the situation 
where the X-ray flux increases by a factor of up to 24 on a time scale 
of seconds.  We discuss the flares in the context of observations and 
theory of Galactic BHCs and compare the flares to those detected from 
Sgr~A*, the super-massive black hole at the Galactic center.  We also 
compare the flares to X-ray bursts that are seen in neutron star systems.  
While some of the flare light curves are similar to those of neutron 
star bursts, the flares have non-thermal energy spectra in contrast to 
the blackbody spectra exhibited in bursts.  This indicates that X-ray 
bursts should not be taken as evidence that a given system contains a 
neutron star unless the presence of a blackbody component in the burst 
spectrum can be demonstrated.

\end{abstract}


] 

\section{Introduction}

X-ray transients provide an opportunity to study accreting compact 
objects over a large range of luminosities and mass accretion rates.
Until recently, most observations of black hole candidate (BHC) 
X-ray transients occurred when the sources were at very high 
luminosity ($\gtrsim 10^{36}$~erg~s$^{-1}$) or very low luminosity
($\lesssim 10^{32}$~erg~s$^{-1}$), and much less observing time 
was devoted to intermediate luminosities (see Chen, Shrader \& Livio 
1997\nocite{csl97} for a review of observations of X-ray transients).
This situation has improved somewhat through observations with the 
{\em Rossi X-ray Timing Explorer} \citep{brs93} and also with imaging 
X-ray observatories.  Over the past few years, we have been observing 
BHC transients during outburst decay with {\em RXTE}, the {\em Chandra 
X-ray Observatory} \citep{weisskopf02}, and also in the radio band 
(see, e.g., Tomsick \& Kaaret 2000\nocite{tk00}; 
Corbel et al.~2001\nocite{corbel01}; Tomsick, Corbel \& 
Kaaret 2001\nocite{tck01}; Kalemci 2002\nocite{kalemci_thesis}).  One 
motivation for these observations is performing detailed studies of 
the transitions between the high-soft (or soft) and low-hard (or hard) 
spectral states \citep{vdk95} where significant changes in the X-ray 
spectral and timing properties occur.  While these transitions 
typically occur at luminosities of about $10^{36}$ erg~s$^{-1}$ 
\citep{nowak95}, our observing program is designed to extend to much 
lower luminosities as we have recognized that new phenomena associated 
with intermediate accretion rates may be found by observing in this 
rather poorly studied luminosity regime.

In the hard state for BHCs, it is likely that the X-ray emission is due 
to inverse Comptonization of soft photons by energetic electrons, but 
there is uncertainty about the system geometry and the mechanism for 
transferring energy to the electrons.  One possibility is that the 
inner edge of the accretion disk recedes from the compact object 
in the hard state, leaving a hot, quasi-spherical, optically thin 
region in the inner portion of the accretion disk.  Another possibility 
that implies a different site for hard X-ray production and a different 
accretion geometry is that the hard X-ray emission is due to a large
number of discrete regions above the accretion disk where magnetic
reconnection events occur \citep{grv79}.  While both of these
possibilities describe physical processes that may occur in BHCs, 
neither provides a complete picture of accreting BHCs as radio 
observations in the hard state indicate the presence of a powerful
compact jet \citep{fender01}.  The compact jet has also been suggested
as a possible contributor to the X-ray emission via inverse 
Comptonization or possibly a synchrotron mechanism 
\citep{mff01,markoff03}.

The BHC transient XTE J1650--500 was discovered in 2001 September
when it was detected in the X-ray band in outburst by the {\em RXTE} 
All-Sky Monitor \citep[ASM,][]{remillard01}.  Pointed {\em RXTE}
observations indicated X-ray spectral and timing properties 
typical of black hole systems:  No X-ray pulsations were detected; 
the energy spectrum was consistent with a combination of a soft 
component, likely from the accretion disk, and a power-law 
component; and band-limited noise and quasi-periodic oscillations
(QPOs) were observed \citep{mss01,rs01,wml01}.  XTE J1650--500
was identified in the optical \citep{ct01} and also the radio
\citep{gum01}, but the radio source was not resolved.  
XTE J1650--500 was observed in the X-ray band with {\em XMM-Newton} 
and {\em Chandra} during its outburst.  The {\em XMM-Newton} 
spectrum represents one of the best examples of a smeared and
gravitationally red-shifted iron K$\alpha$ emission line 
that has been detected for a stellar mass BHC \citep{miller02}.
More recently, \cite{homan03} reported the detection by {\em RXTE} 
of a 250~Hz QPO, making XTE J1650--500 the sixth BHC where
high frequency ($>$40~Hz) QPOs have been detected.  Finally, 
optical observations near quiescence indicate a binary orbital 
period of 5.1~hr and an optical mass function of 
$0.64\pm 0.03$~\Msun~\citep{sf02}.  The value for the mass
function is relatively low for a black hole system and could 
indicate that the system contains a relatively low mass black 
hole, has a low binary inclination ($i<40^{\circ}$), or both.

During the 2001 X-ray outburst from XTE J1650--500, the source 
underwent dramatic changes in its X-ray spectral and timing 
properties.  Generally, the changes were not unusual for
BHC X-ray transients.  The source started off in a hard state, 
with a hard energy spectrum and a high level of timing noise, 
made a transition to a soft spectral state, and then made a 
transition back to the hard state \citep{kalemci03,homan03}.  
Here, we focus on {\em RXTE} observations that were made well 
after the state transition when the source flux was between 10 
and 100 times below the transition level.  In \S 2, we describe 
the observations and our analysis of the data.  We present the 
results of the analysis in \S 3.  The results include the 
discovery of X-ray oscillations and flares, and we previously 
reported the oscillations in an IAU Circular \citep{tomsick_iauc02}.  
We discuss the results in \S 4 and summarize our conclusions 
in \S 5.

\section{Observations and Analysis}

\begin{figure}[t]
\plotone{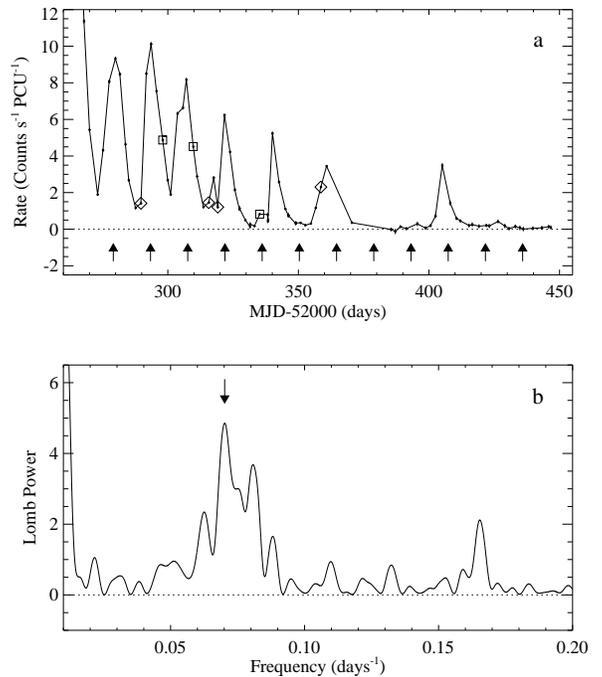}
\vspace{0.2cm}
\caption{(a) The 2.8-20~keV PCA light curve at the end of the 2001 
XTE J1650--500 outburst.  The count rate is the background subtracted, 
top-layer rate for PCU 2.  The diamonds mark observations during which
X-ray flares occurred, and the squares mark the observations where 
simultaneous {\em Chandra} data were obtained.  The rates shown 
include the counts from the X-ray flares.  The arrows are spaced 
by 14.2 days.  (b) The Lomb periodogram for the PCA light curve.  The 
arrow marks the most significant peak in the periodogram, which is at 
14.2 days, but it is not a periodic signal.\label{fig:lc}}
\end{figure}

We obtained 97 {\em RXTE} monitoring observations of 
XTE J1650--500 covering the end of its 2001-2002 outburst.
These ``Target of Opportunity'' observations started on
2001 November 5 (MJD 52,218) and were triggered by the
gradual drop in the X-ray flux that was observed by the 
{\em RXTE}/ASM.  Of the 97 observations, 15 occurred prior 
to a 28 day gap caused by pointing restrictions related to 
the source position relative to the sun.  \cite{kalemci03} 
report on timing analysis of these pre-gap observations, 
during which the 2.8-20 keV source rate was between 69 and 
183 s$^{-1}$ per PCU (Proportional Counter Unit), including 
counts from the top anode layer only.  Here, we focus on the 
82 post-gap {\em RXTE} observations that were made over a 
period of 182 days between 2001 December 22 and 2002 June 22.  
Immediately after the gap, the source rate had dropped to 
21 s$^{-1}$, and for the remaining observations, the source 
rates averaged over the duration of each observation are 
shown in Figure~\ref{fig:lc}a.  The {\em RXTE} observations 
occurred approximately every 2 days with the exception of 
the time period between MJD 52,361 and MJD 52,385, during 
which there was only 1 observation.  The exposure times for 
the 82 post-gap observations were between 384 and 3360~s 
with an average exposure time of 1516~s per observation.  
Long time scale ($\sim$14 day) and large-amplitude X-ray 
oscillations are apparent in the Figure~\ref{fig:lc}a light 
curve.

For the light curve shown in Figure~\ref{fig:lc}a and for 
much of the work described below, we used ``Standard 2'' 
PCA (Proportional Counter Array) data with 16~s time 
resolution and 129 energy channels.  We extracted light 
curves and energy spectra using scripts developed at UC 
San Diego and the University of T\"{u}bingen that 
incorporate the standard software for {\em RXTE} data 
reduction (FTOOLS v5.2).  We used the most recent 
(2002 February) release of the ``Faint Source'' model for
background subtraction.  Obtaining the best possible 
background subtraction is important for some aspects of
this work because of the low source count rate.  To 
achieve this, we used Standard 2 data from the top anode 
layer only and excluded data from PCU 0, which has a higher 
background level due to the loss of its propane layer\footnote{See
http://lheawww.gsfc.nasa.gov/users/craigm/pca-bkg/bkg-users.html
for a detailed analysis of the background model performance.}.
The light curve shown in Figure~\ref{fig:lc}a indicates that
we obtain high quality background subtraction.  For the
final 8 observations, we detect little or no flux from
XTE J1650--500, and the measured flux levels are consistent
with the expected level from the Galactic ridge emission
\citep{vm98} in the PCA field of view for XTE J1650--500.  
Finally, we note that for some of the results described 
below, we used a second data mode with 125~$\mu$s time 
resolution that was obtained simultaneously with the 16~s 
time resolution data.

During 3 of the {\em RXTE} observations, we also obtained 
simultaneous observations with {\em Chandra} (on MJD 52,298, 
MJD 52,309 and MJD 52,335), and these are indicated in 
Figure~\ref{fig:lc}a.  Although we leave a detailed analysis 
of the {\em Chandra} data to a future paper, we refer to 
these observations in \S 3.3 and \S 3.5.

\section{Results}

\subsection{Discovery of X-Ray Flares}

Through an inspection of the 16~s 2.8-20~keV PCA light curves 
for XTE J1650--500, we found that bright and isolated X-ray 
flares occurred during several of the {\em RXTE} observations.
As shown in Figure~\ref{fig:flares}, some of these flares show 
count rate increases by more than an order of magnitude over
the persistent flux and durations near 100~s.  Even though 
these flares are dramatic bursts of X-ray emission, significant 
variability is present in many of the other 16~s light curves.
Thus, to ensure that we find the most extreme flares, we carried 
out a quantitative search for flares using the 16~s light curves.
First, we produced a count rate histogram (i.e., the number of 
light curve time bins in various count rate ranges vs.~count rate) 
for each observation.  Then, we performed a least squares fit to 
these histograms with a Gaussian function.  We obtained three 
parameters from the fit: The Gaussian mean ($r_{mean}$); the 
width ($\sigma$); and the amplitude.  For each observation, we 
calculated $x = (r_{max}-r_{mean})/\sigma$, where $r_{max}$ 
corresponds to the maximum count rate in the 16~s light curve 
for that observation.  Based on the number of observations and 
their duration, if the light curves contain only pure Gaussian 
noise, one expects approximately 1 observation with $x > 4$ and 
no observations with $x > 5$.  In fact, for XTE J1650--500, 8 
(out of 82) observations have $x > 7$, indicating the presence 
of a non-Gaussian component to the variability.  It should be 
noted that, for light curves in general, a large value of $x$ 
does not necessarily indicate the presence of flares because 
a long-term trend could also produce a large value of $x$.  
However, an inspection of the light curves for the 8 XTE 
J1650--500 observations with $x > 7$ indicates that these light 
curves contain large and rapid increases in flux (i.e., flares).  
In this work, we focus on the 5 observations that contain the 
most extreme flares with values of $x$ between 10 and 20.

Closer examination of the flares in the 5 observations indicates 
that the properties of the flare that occurred during the 
observation on 2002 January 9 (MJD 52,283) are significantly 
different from those of the other flares.  The Jan.~9 flare is 
extremely hard relative to the other flares.  While the peak 
rate in the 2.8-20~keV band is similar for the flares in all 
5 observations, the peak 20-60~keV count rate for the Jan.~9 
flare is $18\pm 1$ s$^{-1}$ per PCU compared to non-detections
in this energy band with upper limits of 1-2 s$^{-1}$ per PCU 
for the other flares.  In fact, there is good evidence that the 
Jan.~9 flare did not originate from any source within the 
{\em RXTE} field-of-view (FOV).  Based on the PCA rates observed 
at the peak of the Jan.~9 flare, a source in the FOV should have 
been strongly detected with a count rate of approximately 
600 s$^{-1}$ in HEXTE (High-Energy X-ray Timing Experiment), 
which is the hard X-ray (20-200~keV) instrument that is co-aligned 
with the PCA on-board {\em RXTE} \citep{rothschild98}.  We 
examined the HEXTE light curves for the Jan.~9 observation, and 
did not find any evidence for a count rate increase.  

The properties of the Jan.~9 flare are very similar to 3 flares 
that were detected with {\em RXTE} during observations of other 
sources in 1998.  The 1998 events have been attributed to X-rays 
and gamma-rays produced during solar flares\footnote{http://lheawww.gsfc.nasa.gov/users/keith/non\_cosmic\_bursts/burst\_doc.html}.
To check the hypothesis that the Jan.~9 flare has a solar 
origin and that the other flares we detect during the
XTE J1650--500 observations do not, we obtained publicly 
available X-ray light curves with 5 minute time resolution 
from one of the Geostationary Operational Environmental 
Satellites (GOES-12), which carries the Solar X-ray Imager 
(SXI).  Very bright solar flares were detected with SXI at 
the time of the 2002 Jan.~9 flare and also at the times of 
the 1998 events.  However, solar flares did not occur at the 
times of the other XTE J1650--500 flares, and we conclude 
that the Jan.~9 flare has a solar origin, but the other 
flares do not.  Figure~\ref{fig:flares} displays PCA light 
curves with 4~s time resolution for the 4 XTE J1650--500 
observations (henceforth, observations a-d) where we detect 
non-solar flares.  We detect 6 flares (labeled 1-6 in 
Figure~\ref{fig:flares}) in these observations and report 
on their detailed properties below.

\subsection{Basic X-Ray Flare Properties}

\begin{figure}[t]
\plotone{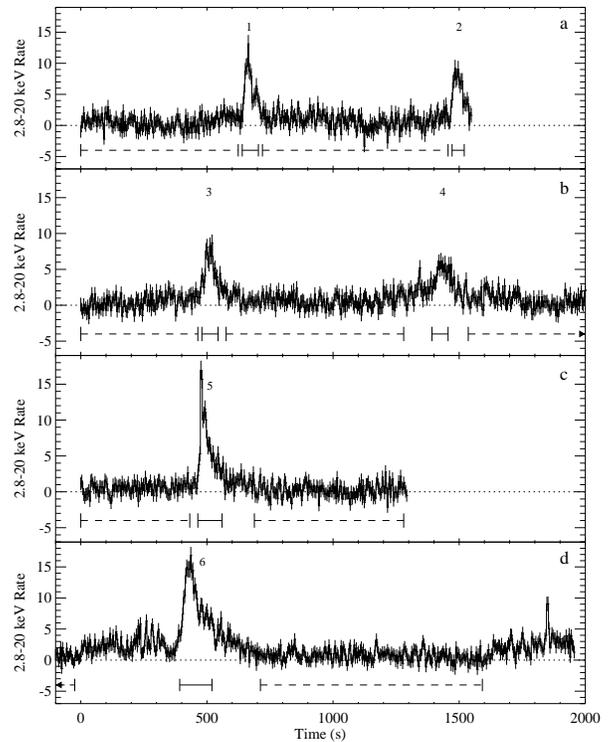}
\vspace{0.2cm}
\caption{The 2.8-20 keV PCA light curves for the 4 observations 
where X-ray flares occurred.  The time resolution is 4~s, and the 
light curves are background subtracted.  To maximize the statistics, 
data from all PCUs that were on, including PCU 0, and all anode 
layers were used, but the rates are divided by the number of 
active PCUs to give counts s$^{-1}$ PCU$^{-1}$.  The flares are 
labeled 1-6, and the flare and non-flare times used for spectral
analysis are marked with solid and dashed lines, respectively, 
below the light curves.  The dates of the observations are (a) 
MJD 52289.7, (b) MJD 52315.7, (c) MJD 52319.1, and (d) MJD 52358.6.
\label{fig:flares}}
\end{figure}

To characterize the flares, we extracted light curves for 
observations a-d with 1~s time resolution.  Although PCA 
background estimates are only provided with 16~s time 
resolution, we used an interpolation procedure to background 
subtract the 1~s light curves.  For each flare, we fitted 
the rise and decay with a function consisting of an 
exponential plus a constant.  In most cases, an exponential 
does not provide a good description of the light curve if 
the peak of the flare is included, but acceptable fits are 
obtained when we include the portion of the light curve up 
to 85\% of the flare peak.  For the 12 fits (rises and decays 
for 6 flares), we obtain reduced $\chi^2$ values between 0.81 
and 1.54 with a mean reduced $\chi^2$ of 1.07, typically for 
110 degrees of freedom.  Table~\ref{tab:basic} gives the 
e-folding rise and decay times for the 6 flares and also the 
flare durations.  The durations are calculated using the 
exponential fits and defining the start of the flare to 
be the time when the rate rose to 20\% above the constant 
level and the end of the flare to be the time when the
rate reached 20\% above the constant level during the 
decay.  The results indicate that the flares have rise 
times of 3-18~s, decay times of 15-98~s, and durations 
of 62-215~s.  For each flare, Table~\ref{tab:basic} also 
includes the ratio of the peak 2.8-20~keV count rate in 
the 4~s light curves to the non-flaring (i.e., persistent) 
rate in the same band.  The non-flaring light curve regions 
are marked in Figure~\ref{fig:flares}.  The peak rates are 
5-24 times higher than the persistent rates.

\subsection{Flare and Non-Flare Energy Spectra}

We studied the spectral properties for observations a-d, 
by producing 2.8-20~keV PCA energy spectra using data 
from the flare and non-flare times shown in 
Figure~\ref{fig:flares}.  We produced energy spectra for 
each flare individually and also a combined spectrum with 
a 464~s integration time.  We determined the non-flare 
times by inspection of the 16~s light curves and included 
all long, continuous segments where the count rate is 
relatively constant.  We produced spectra for each of 
the 4 observations, and the exposure times for the 
non-flare spectra are given in Table~\ref{tab:spectra}.  
For spectral fitting, we used the XSPEC 11.2 software 
package.  Based on fits to PCA energy spectra from 
observations of the Crab nebula that were contemporaneous 
with the XTE J1650--500 observations (see Tomsick, Corbel 
\& Kaaret 2001\nocite{tck01} for a more detailed 
description of this procedure), we included 0.6\% and 
0.3\% systematic errors for energy bins below and above 
8~keV, respectively.  

We fitted the 464~s flare spectrum with power-law and 
blackbody spectral models to determine if the flare emission 
mechanism is non-thermal or thermal.  For these fits, we 
included interstellar absorption using the ``phabs'' model 
and left the column density ($N_{\rm H}$) as a free parameter.  
Using the power-law model results in a very good fit with 
$\chi^{2}/\nu = 23/37$, while the result for the blackbody 
model is a poor fit with $\chi^{2}/\nu = 120/37$.  Thus, 
it is clear that the flare spectrum is not dominated by a
blackbody component.  Furthermore, a two-component power-law 
plus blackbody fit does not provide a significant improvement 
over the power-law alone, and the upper limit on the blackbody 
contribution to the two-component fit is 7\% of the total 
2.8-20~keV flux. 

The non-flare spectra are also well-described by a power-law 
model.  However, for observations a and d, positive 
residuals appear between 6 and 7~keV that may indicate
the presence of an iron K$\alpha$ emission line.  For 
observation d, where the largest residuals occur, the line 
is consistent with being narrow and the line energy is 
$6.6\pm 0.3$~keV.  The equivalent width (EW) is poorly 
constrained to be between 200~eV and 1.4~keV (90\%
confidence).  Based on the improvement in the quality 
of the fit when the line is included in the model, an 
F-test indicates that the line is required at only the
99.4\% level.  For the other non-flare observations and 
for the 464~s flare spectrum the 90\% confidence EW upper 
limits are a few hundred eV or greater -- consistent 
with the EW for observation d.  Thus, for the spectral 
fits described below, we use a model consisting of a 
power-law and a narrow iron line.

For the flare and non-flare spectra, the power-law plus 
line spectral fits indicate that the constraints on the 
column density are poor, and we obtain 90\% confidence 
upper limits on $N_{\rm H}$ in the range 
(2-10)$\times 10^{22}$ cm$^{-2}$.  Our {\em Chandra}
spectra (see \S 2) indicate a column density of about 
$N_{\rm H} = 6\times 10^{21}$ cm$^{-2}$, which is 
consistent with the PCA upper limits, the expected
interstellar value \citep{dl90}, and also the values 
inferred from X-ray and optical observations obtained 
during outburst (Miller et al.~2002\nocite{miller02}; 
Augusteijn et al.~2001\nocite{acg01}; A. Castro-Tirado, 
private communication).  We refitted the spectra after 
fixing $N_{\rm H}$ to the value we obtained from 
{\em Chandra}, and the results for the combined flare 
spectrum and the non-flare spectra are given in 
Table~\ref{tab:spectra}.  While the combined flare 
spectrum does have a smaller best fit value for the 
power-law photon index ($\Gamma$) than the non-flare 
spectra, it is only marginally harder than the non-flare 
spectra for observations b and d.  However, at 
$\Gamma = 1.59\pm 0.08$ (90\% confidence errors), the 
flare spectrum is significantly harder than the spectra 
of the two non-flare observations with the lowest flux, 
a and c, for which $\Gamma = 2.20^{+0.31}_{-0.26}$ and 
$\Gamma = 2.50^{+0.48}_{-0.43}$, respectively.  This 
indicates that significant softening occurs at low flux 
levels, but it is not clear if this is related to the 
flaring behavior.  We refitted the combined flare spectrum 
after subtracting the average non-flare spectrum.  This 
results in a slightly lower (harder) value for $\Gamma$, 
but the change in $\Gamma$ is less than 0.1.

We also fitted the spectra for the individual flares and 
the values of $\Gamma$ obtained are given in 
Table~\ref{tab:basic} along with the fluences for each 
flare.  To determine the fluences, we first summed the 
total number of counts over the duration of the flare 
and subtracted the contribution from the persistent flux.
We determined the start and end times for each flare
as described in \S 3.2.  We then used the spectral shape 
from fitting the individual flares to convert the total 
counts to a fluence in erg~cm$^{-2}$. 

\subsection{Evolution of X-Ray Flare Properties}

\begin{figure}[t]
\plotone{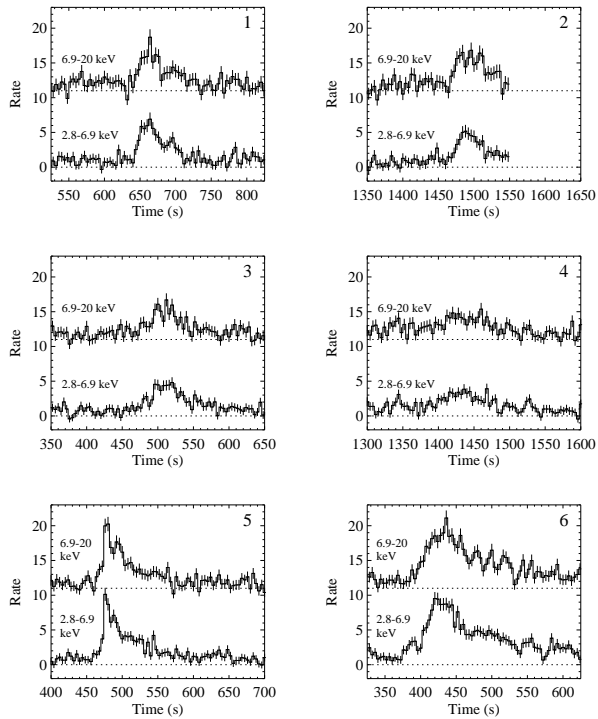}
\vspace{0.2cm}
\caption{Light curves in 2 energy bands for the 6 X-ray flares.
The time resolution is 4~s, and the light curves are background 
subtracted.  To maximize the statistics, data from all PCUs that 
were on, including PCU 0, and all anode layers were used, but the 
rates are divided by the number of active PCUs to give counts 
s$^{-1}$ PCU$^{-1}$.  For each X-ray flare, the soft 2.8-6.9 keV 
band light curve is shown on the bottom.  The hard 6.9-20 keV band 
light curve is shown on top and has been offset by 10 
counts s$^{-1}$ for clarity.\label{fig:evolution}}
\end{figure}

Figure~\ref{fig:evolution} shows the light curves for the 6 
flares in the hard (6.9-20 keV) and soft (2.8-6.9 keV) X-ray
bands.  While there are some specific light curve features 
that appear for only one of the energy bands, the main result
is that there is good overall correlation between the soft 
and hard light curves.  We also examined the hardness ratio 
vs. time using the hard and soft light curves, but the only 
obvious trend is that the flares are somewhat harder than the 
persistent emission, which is consistent with the spectral 
results reported in \S 3.3.  For flare 6 especially, it is
apparent that some of the specific light curve features are
present for only one of the energy bands.  During the rise
of flare 6, near time = 400~s, there is an 8~s dip that only
appears in the soft band light curve.  Also, during the 
flare 6 decay (time = 465-540~s), oscillations with a 
$\sim$20~s time scale are present in the hard band light 
curve only.  In addition, flares 2 and 3 show some evidence
for variability in the hard band that is not present in the
soft band.

\subsection{Long Time Scale Oscillations Observed with {\em RXTE} 
and {\em Chandra}}

Here, we focus on the long time scale and large amplitude X-ray
oscillations seen in Figure~\ref{fig:lc}a.  To determine the 
characteristic frequencies of the oscillation and to search for 
periodicities, we produced a Lomb periodogram \citep{press92} 
of the time series, and it is shown in Figure~\ref{fig:lc}b.
The highest amplitude peak in the periodogram is at a frequency
of 0.070 days$^{-1}$ (14.2 day period), but it is about a 
factor of 2 below the 3-$\sigma$ detection limit for a 
periodic signal.  The fact that there is excess power from 
0.06 to 0.09 days$^{-1}$ suggests the presence of an aperiodic 
signal, and inspection of the light curve agrees with this conclusion.  
In Figure~\ref{fig:lc}a, the arrows are spaced by 14.2 days, and 
while the first 4 peaks are consistent with this period, the 
agreement does not continue after MJD 52,330. 

As the PCA is not an imaging instrument and has a relatively large 
field-of-view ($1^{\circ}$ radius), it is necessary to consider 
whether another source besides XTE J1650--500 could be responsible 
for the phenomena detected by the PCA.  Thus, it is important that
the {\em Chandra} observations mentioned in \S 2 firmly establish 
that XTE J1650--500 is the source of the oscillations.  For all
three {\em Chandra} observations, a source is strongly detected
at the position of the XTE J1650--500 radio counterpart
\citep{gum01}.  The {\em Chandra} spectra are well-described
by a power-law with interstellar absorption, and by fitting
the spectra, we derive 0.5-10~keV fluxes of $5\times 10^{-11}$ 
erg~cm$^{-2}$~s$^{-1}$ for the first two observations and 
$6\times 10^{-12}$ erg~cm$^{-2}$~s$^{-1}$ for the third 
observation.  Comparing these fluxes to the PCA count rates
shown in Figure~\ref{fig:lc}a indicates that the flux levels
measured by {\em Chandra} and the PCA are well-correlated.
As the {\em Chandra} observations prove that XTE J1650--500
was still active when the flares were observed, it is very 
unlikely that the flares come from another source.  Still,
we used the SIMBAD database to search for X-ray sources in 
the PCA FOV, which is a $1^{\circ}$ circle centered on the 
XTE J1650--500 position ($l = 336.70^{\circ}, b = -3.44^{\circ}$).  
We did not find any sources, such as X-ray binaries, that are 
likely to contribute significantly to the X-ray flux detected 
by the PCA.  Further, the non-thermal flare spectrum rules out 
the possibility that the flares come from an X-ray burster in 
the PCA FOV (see \S 4.5 for more discussion of X-ray bursts).

\section{Discussion}

During {\em RXTE} observations of the BHC XTE J1650--500, we 
have discovered 6 X-ray flares lasting from 62-215~s and 
peaking at count rate levels between 5 and 24 times the 
persistent rates.  The flares occurred at the end of the first 
and currently only X-ray outburst that has been detected from 
this X-ray transient when the persistent flux was a factor of 
$10^{3}$ below the the peak flux level for the outburst.  
Although strong X-ray variability is commonly observed for 
black hole X-ray binaries, the isolated X-ray flares that we 
observe are unusual.  In addition, we find long term aperiodic 
oscillations with a characteristic time scale of 14.2 days.  
Below, we discuss the implications of the XTE J1650--500 
oscillations and X-ray flares by comparing them to phenomena 
that have been seen for other sources, including long and short 
time scale variability in black hole binaries, the X-ray flares 
that have been detected from Sgr~A*, the super-massive black 
hole at the center of the Galaxy, and X-ray bursts in neutron 
star binaries.

\subsection{XTE J1650--500 Oscillations}

The fact that the $\sim$14 day X-ray oscillations are an aperiodic 
phenomenon likely rules out the possibility that they are 
directly related to disk precession, and they are obviously not 
orbital since it is now known that XTE J1650--500 has an orbital 
period of 5.1~hr \citep{sf02}.  However, oscillations have previously 
been observed in the optical at the ends of outbursts for a number 
of short orbital period BHC transients as well as dwarf novae 
\citep{kuulkers98}.  The general properties of the optical 
oscillations can be explained in the framework of disk instability 
models (DIM) for dwarf novae \citep{ost97,hlw00} and possibly
for X-ray transients \citep{menou00}.  For DIM, the level of 
emission increases when the disk becomes hot enough for the 
hydrogen in the disk to be ionized.  This leads to a higher 
viscosity and an increase in the mass accretion rate onto the 
compact object.  Oscillations can be produced by the propagation 
of heating and then cooling waves through the disk \citep{hlw00}.
For BHC systems, the best examples of these optical oscillations 
(also called ``mini-outbursts'') include GRO J0422+32 \citep{ct93,ci95}, 
GRS 1009--45 \citep{bo95}, and, more recently, XTE J1859+226 
\citep{zurita02}.  For XTE J1650--500, optical oscillations occurred 
contemporaneously with the X-ray oscillations we are reporting in this 
work (C. Bailyn, private communication).  The optical oscillations 
have similar characteristic time scales of 10-100 days for all 4 
sources.  While X-ray coverage during optical oscillations was not 
obtained for GRS 1009--45 and was sparse for GRO J0422+32 and 
XTE J1859+226, the X-ray observations obtained for the latter 2 
sources showed that they were active in the X-ray band during the 
oscillations \citep{shrader97,th00}.  Another short orbital period 
BHC transient, 4U 1543--47, showed large variations in the X-ray 
band in a poorly sampled light curve obtained at the end of its 
1971-1972 outburst \citep{csl97}.  Overall, the similarities 
between the XTE J1650--500 oscillations and those that have been 
seen for other short orbital period BHC transients suggests that 
they are related phenomena.  If this is the case, our observations 
of XTE J1650--500 represent the first time that good X-ray coverage 
has been obtained during a series of mini-outbursts.

\subsection{XTE J1650--500 X-Ray Flares and BHC Variability}

While it is likely that the XTE J1650--500 oscillations are similar
to phenomena that have been observed before (mostly in the optical
band) for BHCs, it is not clear that the same can be said for the 
X-ray flares.  In certain spectral states, BHC light curves certainly 
do exhibit a high level of variability, and their X-ray light curves 
contain flares that have inspired ``shot noise'' models \citep{terrell72}.  
However, in these models, it is the combination of a very large number 
of shots that can lead to the overall variability observed in BHCs, 
which can be as fast as millisecond time scales.  The shots, and 
also the flares that are typically observed in BHC light curves, 
have much smaller amplitudes relative to the average flux levels 
and much shorter time scales than the XTE J1650--500 flares (see 
Miyamoto et al.~1992\nocite{miyamoto92} for examples of light 
curves for other BHCs).  

We found one example in the literature of a black hole system
that exhibited a series of X-ray flares with some similarities
to the XTE J1650--500 flares.  V4641~Sgr is an unusual X-ray 
transient that was discovered in 1999 February by {\em RXTE} 
and {\em BeppoSAX}.  It is often referred to as a ``fast'' 
transient because it had an outburst in 1999 September that 
only lasted for about a day rather than the typical several 
months.  The source showed extreme X-ray variability when it 
was bright, and, at the end of its most active period in 1999 
September, it exhibited relatively isolated X-ray flares 
\citep{wv00}.  Three or four X-ray flares occurred in a 
200~s light curve shown in Figure~1c of \cite{wv00}.  The 
X-ray flares had rise times $<$10~s, durations of 20-50~s, 
flux increases by factors of 10-50 over the non-flaring flux, 
and roughly exponential decays at the ends of the flares.  
The similarities of the flare parameters and the fact that the 
flares for both V4641~Sgr and XTE J1650--500 occurred when the 
non-flaring flux was several orders of magnitude below the 
peak outburst flux for the respective sources leads us to 
suggest the possibility that the flares from the two sources 
could be related.

Although we do not believe that similar X-ray flares have
been seen before in BHC sources other than XTE J1650--500 and
possibly V4641~Sgr, this could partly be a selection effect.
As discussed in \S 1, BHCs have been most often observed in 
outburst at high luminosity or in quiescence.  From the 
optical observations of XTE J1650--500 that have been made 
\citep{sf02,acg01,gw02}, we estimate that the source distance 
is between 2 and 6~kpc, and here and below we use a fiducial 
value of 4~kpc for luminosity estimates.  Thus, the typical 
non-flare X-ray flux of $1.7\times 10^{-11}$ erg~cm$^{-2}$~s$^{-1}$
(2.8-20~keV, unabsorbed) corresponds to an luminosity of 
$3\times 10^{34} (d/4~{\rm kpc})^{2}$ erg~s$^{-1}$, which is 
in a luminosity regime where BHCs have not been as well-studied.  
Thus, we cannot rule out the possibility that X-ray flares
commonly occur for BHCs at intermediate luminosities.

The fact that multiple X-ray flares are detected from 
XTE J1650--500 during observations a and b, suggests the
possibility that XTE J1650--500 entered into a state (or
perhaps an accretion regime) that favors flare production.
It is interesting that the flares for observations a and b
are both separated by 800-900~s, and this time scale may
provide a clue about the flare emission mechanism.  
The observation d light curve also supports the idea that 
XTE J1650--500 enters into a flaring state since a short 
flare is seen at the end of that observation (near time = 
1850~s).  Furthermore, the {\em RXTE} observation following 
observation d (on MJD 52361) is one of the observations for
which higher variability was detected, and short flares
are also detected during that observation.  Finally, a 
higher level of variability was detected for the {\em RXTE} 
observation that occurred between observations b and c on 
MJD 52318.

\subsection{X-Ray Flares and Black Hole Accretion Theory}

The XTE J1650--500 X-ray flares demonstrate that physical models 
for BHC systems must be able to account for the situation where 
the X-ray flux increases by a factor of up to 24 on a time scale 
on a time scale of seconds (based on flare 5).  While this time 
scale does not provide a useful model-independent constraint on the 
size of the emission region because the physical size of the system 
(i.e., the binary separation) is only a few light-seconds, we have 
considered whether the flares have implications for certain 
accretion scenarios.  For flares produced by a change in mass 
accretion rate, the flux is expected to change on the viscous 
time scale in the X-ray emitting region of the accretion disk.  
For thin ($H\ll R$) and thick ($H\sim R$) accretion disks, where 
$H$ is the accretion disk thickness and $R$ is the radial distance 
from the compact object, the viscous time scale is given by 
$t_{v} = \alpha^{-1} (R/H)^{2}~t_{d}$, where $\alpha$ is a 
dimensionless parameter that is proportional to the viscosity 
and $t_{d}$ is the dynamical time scale \citep{fkr92}.  The
dynamical (i.e., Keplerian) time scale depends on the mass of
the compact object, $M$, and $R = r R_{S}$ ($R_{S} = 2GM/c^{2}$, 
where $G$ is the gravitational constant and $c$ is the speed of 
light) according to $t_{d} = 88\mu$s $(M/\Msun)~r^{3/2}$.
Based on the current distribution of black hole masses, it is
likely that XTE J1650--500 contains a black hole with a mass
of 5-15\Msun, and we assume a mass of 10\Msun.  In the extreme
case of $\alpha = 1$ \citep{fkr92} and a thick accretion 
disk ($R = H$), $t_{v} = 1$~s corresponds to $r = 110$, while
a more realistic value of $\alpha = 0.1$ gives $r = 23$.  Thus, 
if the flares are due to an increase in the mass accretion rate,
they must originate in a region smaller than the inner 
$\sim$20-100~$R_{S}$ of the accretion disk.  While this does
not strongly constrain theoretical models, the possibility of 
larger hard X-ray emission regions has been previously suggested 
(e.g., Esin, McClintock \& Narayan 1997\nocite{emn97}).  

Some theoretical models for accreting black hole systems predict 
highly variable X-ray emission at low luminosities.  \cite{mf02} 
discuss the possibility of a ``coronal outflow dominated accretion 
disk model,'' where a geometrically thin and optically thick 
accretion disk dissipates much of its gravitational energy in a 
magnetic corona.  This is an attractive model since it incorporates 
angular momentum transport via magnetic fields \citep{bh91} and 
also the presence of outflows such as the compact jets that have 
been observed from accreting black holes at low luminosities 
\citep{fender01}.  In this model, non-thermal X-ray flares can 
be produced through gradual magnetic field generation that can 
lead to a brightening of the corona \citep{mf02}.  In another 
model that has been suggested for low luminosity black hole 
systems, the black hole accretes via a ``hot settling flow'' 
\citep{mm02}.  Here, the black hole is magnetically coupled to
the accretion flow, allowing a rapidly rotating black hole to 
transfer its angular momentum to the accretion flow as in 
\cite{bz77}.  Hard X-ray spectra and high variability are
expected for this accretion mode.  In fact, \cite{mm02} 
specifically predict that XTE J1650--500 may exhibit this 
behavior based on the spectral determination that it contains 
a rapidly rotating black hole \citep{miller02}.

\subsection{Comparison between the X-Ray Flares in XTE J1650--500 and Sgr A*}

Here, we consider the possibility that the XTE J1650--500 flares 
are related to the isolated X-ray flares that have been detected 
from Sgr~A* \citep{baganoff01,goldwurm03}.  Such a comparison is 
interesting because these are both low luminosity BHC or black hole 
systems, and the overall shapes of the flare light curves and 
the evolution of the energy spectra are similar in the two cases.  
For the Sgr~A* flare seen by {\em Chandra} in 2000 October, the 
flux increases by a factor of about 45, and during the 10~ks
flare, the X-ray spectrum hardens and is non-thermal 
\citep{baganoff01}.  While the results for the 2000 October flare 
have been reported in the most detail, subsequent {\em Chandra} 
observations of Sgr~A* yielded the detection of several more 
flares, and these suggest that flux increases by a factor of 
5-25, similar to the XTE J1650--500 flares, are probably more 
typical \citep{baganoff02}.  While the flare durations and rise 
times are about a factor of 100 less for XTE J1650--500 (durations 
of 100~s for XTE J1650--500 compared to 10~ks for Sgr~A* and rise 
times of 10~s for XTE J1650--500 compared to 1~ks for Sgr~A*), 
it is notable that that the ratios for the two quantities
are approximately the same.  For {\em Chandra} observations of 
Sgr~A*, the power-law photon index ($\Gamma$) is 
$2.2^{+0.5}_{-0.7}$ during non-flare times and $1.3^{+0.5}_{-0.6}$ 
during the 2000 October flare \citep{baganoff01}, which is 
consistent with the spectral evolution we observe for XTE J1650--500.  

Although the phenomenology of the flares in Sgr~A* and 
XTE J1650--500 have similarities, we must also consider the 
differences between these systems.  Beyond the fact that the mass 
of the Sgr~A* black hole is $2.6\times 10^{6}$\Msun~\citep{eg97} 
compared to the likely $\sim$10\Msun~black hole in XTE J1650--500, 
the flares we observe for XTE J1650--500 occur when the persistent
Eddington-scaled luminosity is much higher than for Sgr~A*.
For a distance of 4~kpc and a compact object mass of 10\Msun, 
the persistent flux of $10^{-11}$ erg~cm$^{-2}$~s$^{-1}$ for 
XTE J1650--500 (see Table~\ref{tab:spectra}) gives 
$L_{x}/L_{Edd} = 2\times 10^{-5}$.  For Sgr~A*, the persistent 
X-ray luminosity is $2.2\times 10^{33}$ erg~s$^{-1}$ 
\citep{baganoff01}, indicating $L_{x}/L_{Edd} = 5\times 10^{-12}$.
Even if we consider the bolometric luminosity of 
$\sim$$10^{37}$ erg~s$^{-1}$ for Sgr~A* \citep{narayan98}, 
$L/L_{Edd} = 2\times 10^{-8}$, which is three orders of 
magnitude lower than XTE J1650--500.  Thus, we consider
the possibility that the flares in the two systems are 
related with the caveat that their accretion properties 
are likely to be significantly different.

If the flares are produced by similar processes, the fact 
that the flare time scales are longer for Sgr~A* than for 
XTE J1650--500 is not surprising given the much larger black 
hole mass for Sgr~A*.  However, it is not obvious why the
time scales would be different by only a factor of 100.  
For example, dynamical and viscous time scales at the same 
accretion disk radius, $r = R/R_{S}$, are proportional to 
the black hole mass.  Thus, if the flares were produced 
by an increase in the mass accretion rate, the viscous 
time scale would be relevant \citep{lm02}, and a difference 
in time scale by a factor of $3\times 10^{5}$ would be 
predicted.  \cite{baganoff01} discuss the possibility that 
the Sgr~A* flare emission could have a synchrotron or a 
synchrotron self-Compton (SSC) mechanism.  This could 
occur if electrons were accelerated in magnetically active 
regions of an accretion flow (perhaps due to magnetic 
reconnection events).  In these cases, the flare durations 
may be related to the synchrotron or SSC cooling time scales, 
which are proportional to $\gamma^{-1} B^{-2}$ \citep{bbr84}, 
where $\gamma$ is the Lorentz factor for the X-ray producing
electrons and $B$ is the magnetic field strength.  The 
fact that the flare X-ray spectra are similar for Sgr~A* 
and XTE J1650--500 may suggest that $\gamma$ is not 
extremely different for the two systems, but, if $B$ was 
an order of magnitude larger in the XTE J1650--500 emission 
region, then the predicted time scale would be a factor of 
100 lower for this system, as observed.  However, the
calculations of \cite{dcf97} suggest that the difference
in magnetic field strength should be much greater than
an order of magnitude.  Finally, jet-based models have been 
suggested to explain the Sgr~A* flares \citep{mfyb01}.  In
these models, the X-ray emission has a synchrotron or SSC
mechanism, but the electrons are continuously shock 
accelerated so that the cooling time scales do not relate
to the predicted flare duration in a simple way.  More work is 
required to determine if any of these models can explain the 
time scale ratio of 100 between the XTE J1650--500 and Sgr~A* 
flares.

\subsection{X-Ray Flares and X-Ray Bursts}

Accreting low magnetic field neutron stars exhibit type I X-ray
bursts due to thermonuclear instabilities that result from 
accretion onto the neutron star surface \citep{lvt95}.  While
XTE J1650--500 is a BHC and is not expected to produce type I 
X-ray bursts, the flare rise times, the decay times, and the 
ratio between the peak X-ray flux and the persistent flux are 
all typical of type I X-ray bursts \citep{lvt95}.  Even though 
the light curve shapes for the flares are typical of type I 
X-ray bursts, the spectral evolution is not.  The emission 
mechanism for type I X-ray bursts is thermal, and X-ray bursts 
have energy spectra that are dominated by a blackbody component.  
As shown in \S 3.3, the possibility that the XTE J1650--500 
X-ray flares have a significant blackbody component is strongly 
ruled out.  

Although type II X-ray bursts are not as well understood and 
have only been detected from two sources to date, it is notable 
that some of the XTE J1650--500 flare light curves have properties 
similar to type II bursts.  For flare 6, the variability leading 
up to the flare (time = 0-300~s in Figure~\ref{fig:flares}), the 
dip right before the flare (time near 360~s), and the oscillations 
during the decay have similarities to the light curves of type II
bursts \citep{lvt95,lewin96}.  However, like type I bursts, type
II bursts also exhibit thermal spectra \citep{lvt95,marshall01} 
in contrast to the XTE J1650--500 flares.  If XTE J1650--500 
contains a black hole, then our conclusion that the X-ray flares 
are not type I or type II X-ray bursts indicates that X-ray bursts 
should not be taken as evidence for a neutron star in the system 
unless it can be shown that the burst energy spectrum is dominated 
by a blackbody component.  This conclusion may have implications 
for estimates of the numbers of neutron star binaries in the 
Galaxy if a fraction of the putative X-ray bursters without 
high quality burst spectra actually contain black holes.

While the energy spectrum provides strong evidence against the 
XTE J1650--500 X-ray flares being X-ray bursts, we note that 
this possibility is also disfavored from luminosity 
considerations.  Typically, peak X-ray burst luminosities are 
between $0.1 L_{Edd}$ and $L_{Edd}$.  However, the flux at the 
peak of the XTE J1650--500 outburst is about 100 times higher 
than the peak of the strongest X-ray flare, indicating that if 
the X-ray flares are close to $L_{Edd}$, then the source was 
super-Eddington for most of the outburst, which seems 
implausible.  In addition, given the observed flux, a peak 
X-ray flare luminosity of $0.1 L_{Edd}$ implies a distance 
between 28~kpc and 89~kpc for compact object masses between 
1\Msun~and 10\Msun, and these large distances are ruled-out
by the optical detections of the source near quiescence.  However, 
these luminosity arguments may not be absolute proof against 
X-ray bursts because there are some unusual cases where bursts 
in other systems have been classified as type I X-ray bursts 
even though they are significantly sub-Eddington (e.g., 
Gotthelf \& Kulkarni 1997\nocite{gk97}; Garcia \& Grindlay 
1987\nocite{gg87}).  On the other hand, for the examples in 
the two preceding references, it has not been clearly 
demonstrated that non-thermal burst energy spectra are 
ruled out.

\section{Summary and Conclusions}

In this work, we report on {\em RXTE} observations of the
BHC transient XTE J1650--500 made at the end of its first, 
and currently only, outburst.  By observing at low 
luminosities, we found new or previously poorly studied X-ray 
phenomena, including bright $\sim$100~s X-ray flares and long 
time scale oscillations of the X-ray flux.  Over several months, 
the X-ray light curve shows strong aperiodic oscillations with 
a characteristic time scale of 14.2 days during which the X-ray 
flux varies by an order of magnitude.  We argue that the 
oscillations are probably related to the optical oscillations 
(also called ``mini-outbursts'') that have been observed for 
other short orbital period BHC transients such as GRO J0422+32, 
GRS 1009--45, and XTE J1859+226.  If this is the case, our 
observations represent the first time that good X-ray coverage 
has been obtained during a series of mini-outbursts.

A uniform analysis of the 16~s {\em RXTE} light curves leads
to the detection of 6 X-ray flares in 4 observations with 
durations between 62 and 215~s and peak fluxes that are 5-24 
times higher than the persistent flux.  The flare energy spectra 
are non-thermal, being well-described by a power-law with a 
photon index of $\Gamma = 1.59\pm 0.08$ (90\% confidence
errors).  The upper limit on a blackbody contribution to the 
flare is 7\% of the 2.8-20~keV flux.  The flares are isolated 
events that occur when the persistent luminosity is low, 
near $3\times 10^{34} (d/4~{\rm kpc})^{2}$ erg~s$^{-1}$
(2.8-20~keV), making them unusual, if not unique, for BHCs.  
The XTE J1650--500 X-ray flares demonstrate that physical models 
for BHC systems must be able to account for the situation where 
the X-ray flux increases by a factor of up to 24 on a time 
scale of seconds (based on flare 5).  We briefly discuss two 
theoretical models for accreting black holes that predict 
highly variable X-ray emission at low luminosities
\citep{mf02,mm02}.  

We also compare the XTE J1650--500 flares to the X-ray flares 
that have recently been observed from Sgr~A* by {\em Chandra}
and {\em XMM-Newton} as these are both accreting black holes
or BHCs that exhibited isolated X-ray flares during periods of 
low persistent luminosity (although we note that $L_{x}/L_{Edd}$ 
and bolometric $L/L_{Edd}$ are considerably lower for Sgr~A*).  
The increases in flux and spectral evolution for the flares in 
the two systems are comparable, while the time scales (rise 
time and duration) are about 100 times faster for XTE J1650--500.  
Given the factor of $\sim$$3\times 10^{5}$ difference in compact 
object mass between the two systems (assuming that XTE J1650--500 
contains a 10\Msun~black hole), it is not surprising that the 
time scales are faster for XTE J1650--500.  However, the simplest 
physical models suggest that the time scale difference should 
be considerably larger.  Overall, we cannot rule out the 
possibility that the flares in the two systems are related, 
and more work is required on this topic.

Finally, we note that some of the XTE J1650--500 flares
have light curves with similarities to type I X-ray bursts
that are seen in neutron star systems, including rise times
of seconds and exponential decays with e-folding times of
tens of seconds.  However, the non-thermal energy spectra
for the flares are in contrast to the spectra of type I
(and also type II) X-ray bursts, which are dominated by 
blackbody emission.  Thus, the XTE J1650--500 flares 
indicate that X-ray bursts should not be taken as evidence 
for the presence of a neutron star in a given system unless 
the presence of a blackbody component can be demonstrated.

\acknowledgements

We would like to thank C. Bailyn for sharing results from
optical observations of XTE J1650--500 ahead of publication.
JAT acknowledges useful discussions with J. Miller, R. Hynes, 
A. Castro-Tirado, W. Yu, and F. Baganoff.  SC acknowledges 
S. Markoff for stimulating discussions.  We acknowledge useful 
comments and suggestions from an anonymous referee.  The SIMBAD 
database and also information from the Space Environment Center, 
Boulder, CO, National Oceanic and Atmospheric Administration 
(NOAA), US Dept. of Commerce were used in preparing this paper.  
We would like to thank all scientists who contributed to the 
T\"{u}bingen Timing Tools.  JAT acknowledges partial support 
from NASA grant NAG5-10886 and Chandra award number GO2-3056X 
issued by the Chandra X-ray Observatory Center, which is
operated by the Smithsonian Astrophysical Observatory for
and on behalf of NASA under contract NAS8-39073.  PK 
acknowledges partial support from NASA grant NAG5-7405.
EK was partially supported by T\"UB\.ITAK.


\begin{table}[h]
\caption{Basic Flare Properties\label{tab:basic}}
\begin{minipage}{\linewidth}
\footnotesize
\begin{tabular}{cccccccc} \hline \hline
Flare & 
MJD\footnote{Modified Julian Date at the peak of the X-ray flare.}--52,000 &
Rise Time\footnote{See text for precise definition.}& 
Decay Time$^{b}$ & 
Duration$^{b}$ & 
Count Rate & 
\\
Number &
(days) &
(s) &
(s) &
(s) &
Increase\footnote{Ratio of the peak count rate in the 4~s light curves to the persistent 2.8-20~keV rate.} &
$\Gamma$\footnote{Power-law photon index for the flare energy spectra with 90\% confidence errors.} &
Fluence\footnote{2.8-20~keV burst fluence in erg cm$^{-2}$ for the duration of the flare.}
\\ \hline
1 & 289.69046 & $7\pm 3$ & $25\pm 6$ & $65\pm 11$ & $12.4\pm 1.4$ & $1.92^{+0.29}_{-0.27}$ & $2.8\times 10^{-9}$\\
2 & 289.70009 & $4.2\pm 1.6$ & $17\pm 4$ & $62\pm 7$ & $8.7\pm 1.3$ & $1.62^{+0.29}_{-0.27}$ & $2.7\times 10^{-9}$\\
3 & 315.65935 & $13\pm 4$ & $15\pm 4$ & $67\pm 9$ & $8.2\pm 1.1$ & $1.51^{+0.29}_{-0.27}$ & $2.5\times 10^{-9}$\\
4 & 315.67009 & $18\pm 10$ & $33\pm 11$ & $118\pm 24$ & $5.8\pm 1.0$ & $1.63^{+0.35}_{-0.30}$ & $2.9\times 10^{-9}$\\
5 & 319.11676 & $2.9\pm 0.6$ & $37\pm 5$ & $75\pm 8$ & $23.5\pm 1.8$ & $1.68^{+0.21}_{-0.20}$ & $3.9\times 10^{-9}$\\
6 & 358.59491 & $18\pm 3$ & $98\pm 8$ & $215\pm 14$ & $15.6\pm 1.3$ & $1.61^{+0.13}_{-0.12}$ & $1.2\times 10^{-8}$\\
\end{tabular}
\end{minipage}
\end{table}

\begin{table}[h]
\caption{Flare and Non-Flare Energy Spectra\label{tab:spectra}}
\begin{minipage}{\linewidth}
\footnotesize
\begin{tabular}{cccccccc} \hline \hline
Obs. & 
F/NF\footnote{Flare (F) or non-flare (NF).} & 
Exposure\footnote{Exposure time included in the spectra.} (s) & 
$\Gamma$\footnote{Power-law photon index for the energy spectra with 90\% confidence errors.} &
Flux\footnote{2.8-20~keV absorbed flux in erg~cm$^{-2}$~s$^{-1}$.}
\\ \hline
a-d & F & 464 & $1.59\pm 0.08$ & $1.09\times 10^{-10}$\\ 
a & NF & 1392 & $2.20^{+0.31}_{-0.26}$ & $1.47\times 10^{-11}$\\
b & NF & 1712 & $1.75^{+0.24}_{-0.22}$ & $1.67\times 10^{-11}$\\
c & NF & 1056 & $2.50^{+0.48}_{-0.43}$ & $9.61\times 10^{-12}$\\
d & NF & 1088 & $1.72^{+0.31}_{-0.28}$ & $1.69\times 10^{-11}$\\
\end{tabular}
\end{minipage}
\end{table}

\end{document}